\documentclass[aps,prl,twocolumn,superscriptaddress,amsmath]{revtex4-2}
\usepackage{graphicx}
\usepackage{hyperref}
\usepackage{mathrsfs}
\usepackage{bm}
\usepackage{color}
\usepackage{siunitx}
\usepackage[capitalize]{cleveref}
\hypersetup{hypertex=true,
	colorlinks=true,
	anchorcolor=blue,
	linkcolor=blue,
    citecolor=blue,
	urlcolor=blue}
\begin{document}

\title{Majorana modes in helical altermagnet without net magnetism and spin-orbit coupling}

\author{Xing-Jian Yi}
\affiliation{International Center for Quantum Materials, School of Physics, Peking University, Beijing 100871, China}
\affiliation{Hefei National Laboratory, Hefei 230088, China}
\author{Yue Mao}
\affiliation{International Center for Quantum Materials, School of Physics, Peking University, Beijing 100871, China}
\author{Cheng-Ming Miao}
\affiliation{International Center for Quantum Materials, School of Physics, Peking University, Beijing 100871, China}
\author{Qing-Feng Sun}
\email[]{sunqf@pku.edu.cn}
\affiliation{International Center for Quantum Materials, School of Physics, Peking University, Beijing 100871, China}
\affiliation{Hefei National Laboratory, Hefei 230088, China}

\date{\today}

\begin{abstract}
We propose a scheme to realize topological superconductor and Majorana bound states (MBSs) in a one-dimensional metal nanowire on the surface of a helical altermagnet and in proximity to an \emph{s}-wave superconductor, removing the requirement of conventional spin-orbit coupling and net magnetism.
Through gauge transformation, we demonstrate that the helical frame naturally induces spin-momentum locking while the altermagnetism breaks time-reversal symmetry.
The topological superconducting phase is well tuned by chemical potential, altermagnet strength, and helical frequency. Besides, our transport calculation results reveal quantized conductance signatures: a $2e^2/h$ zero-bias peak at nanowire ends and a $4e^2/h$ tunneling conductance at the domain wall of nanowires with opposite chirality, detected via metal lead and scanning tunneling microscopy, respectively. Our research offers new perspectives on finding MBSs.
\end{abstract}
\maketitle

\emph{Introduction}.-Majorana bound states (MBSs) \cite{alicea_NonAbelian_2011, leijnse_Introduction_2012, lutchyn_Majorana_2018, stanescu_Majorana_2013, mao_Phase_2025}, exotic quasiparticles that are their own antiparticles, have attracted considerable interest for their non-Abelian statistics \cite{alicea_New_2012, beenakker_Search_2013, ivanov_NonAbelian_2001} and potential applications in fault-tolerant topological quantum computation \cite{dassarma_Topologically_2005, kitaev_Faulttolerant_2003, sarma_Majorana_2015, aghaee_Interferometric_2025}.
A number of schemes to realize MBSs have been put forward \cite{alicea_Majorana_2010, braunecker_Interplay_2013, brydon_Topological_2015, chung_Topological_2011, cook_Majorana_2011, flensberg_Tunneling_2010, fu_Superconducting_2008, fu_Josephson_2009, he_Selective_2014, kim_Helical_2014, kjaergaard_Majorana_2012, klinovaja_Composite_2012, klinovaja_Topological_2013, law_Majorana_2009, sau_Generic_2010, nadj-perge_Proposal_2013, oreg_Helical_2010, lutchyn_Majorana_2010, chen_ChiralityInduced_2024, miao_Topological_2022, zhang_Nonendpoint_2024, zhuang_Anomalous_2022, mao_Charge_2021, liu_ZeroBias_2012}, including topological insulators \cite{cook_Majorana_2011, fu_Superconducting_2008, fu_Josephson_2009}, semiconductors with strong spin-orbit coupling (SOC)\cite{alicea_Majorana_2010, lutchyn_Majorana_2010, oreg_Helical_2010, sau_Generic_2010}, and magnet atom chains \cite{braunecker_Interplay_2013, brydon_Topological_2015, kim_Helical_2014, klinovaja_Topological_2013, nadj-perge_Proposal_2013}.
Meanwhile, there is experimental evidence supporting the existence of MBSs in various systems \cite{das_Zerobias_2012, deng_Anomalous_2012, jeon_Distinguishing_2017, kim_Tailoring_2018, mourik_Signatures_2012, nadj-perge_Observation_2014, sun_Majorana_2016, wang_Evidence_2018, wang_Coexistence_2012, zhang_Observation_2018}, such as magnet atom chains and vortex of topological superconductor (TSC) \cite{wang_Coexistence_2012, zhang_Observation_2018, wang_Evidence_2018, nadj-perge_Observation_2014}.
Generally, MBSs appear in \emph{p}-wave superconducting order \cite{kitaev_Unpaired_2001}. However, inducing \emph{p}-wave superconducting order usually requires the coexistence of
SOC and external magnet field or adjacent ferromagnet, which imposes substantial constraints for practical implementation.
Consequently, realizing MBSs in systems without SOC and net magnetism is imperative.

A promising proposal to realize MBSs is based on a one-dimensional (1D) semiconducting nanowire with strong SOC in proximity to \emph{s}-wave superconductor (SC) under an external magnet field \cite{oreg_Helical_2010, lutchyn_Majorana_2010}.
The realization of MBSs could be simplified if a scheme requiring neither SOC nor external magnet field exists.
Recently, altermagnet, a class of colinear antiferromagnet, has attracted much attention \cite{ fedchenko_Observation_2024, gonzalezbetancourt_Spontaneous_2023, gonzalez-hernandez_Efficient_2021, krempasky_Altermagnetic_2024, li_Majorana_2023, liu_Absence_2024, reimers_Direct_2024, smejkal_Conventional_2022, smejkal_Emerging_2022, song_Altermagnets_2025, yi_Spin_2025, wan_Altermagnetisminduced_2025}. It features compensated antiparallel magnet order in real space with opposite spin-sublattice connected by crystal-rotation symmetry and demonstrates a spin-splitting energy band structure in momentum space.
Combining with the feature of altermagnet, a recent work realizes 1D TSC and MBSs without net magnetism \cite{ghorashi_Altermagnetic_2024}. However, the SOC is still an essential element.

\begin{figure}[b]
	\includegraphics[width=1\columnwidth]{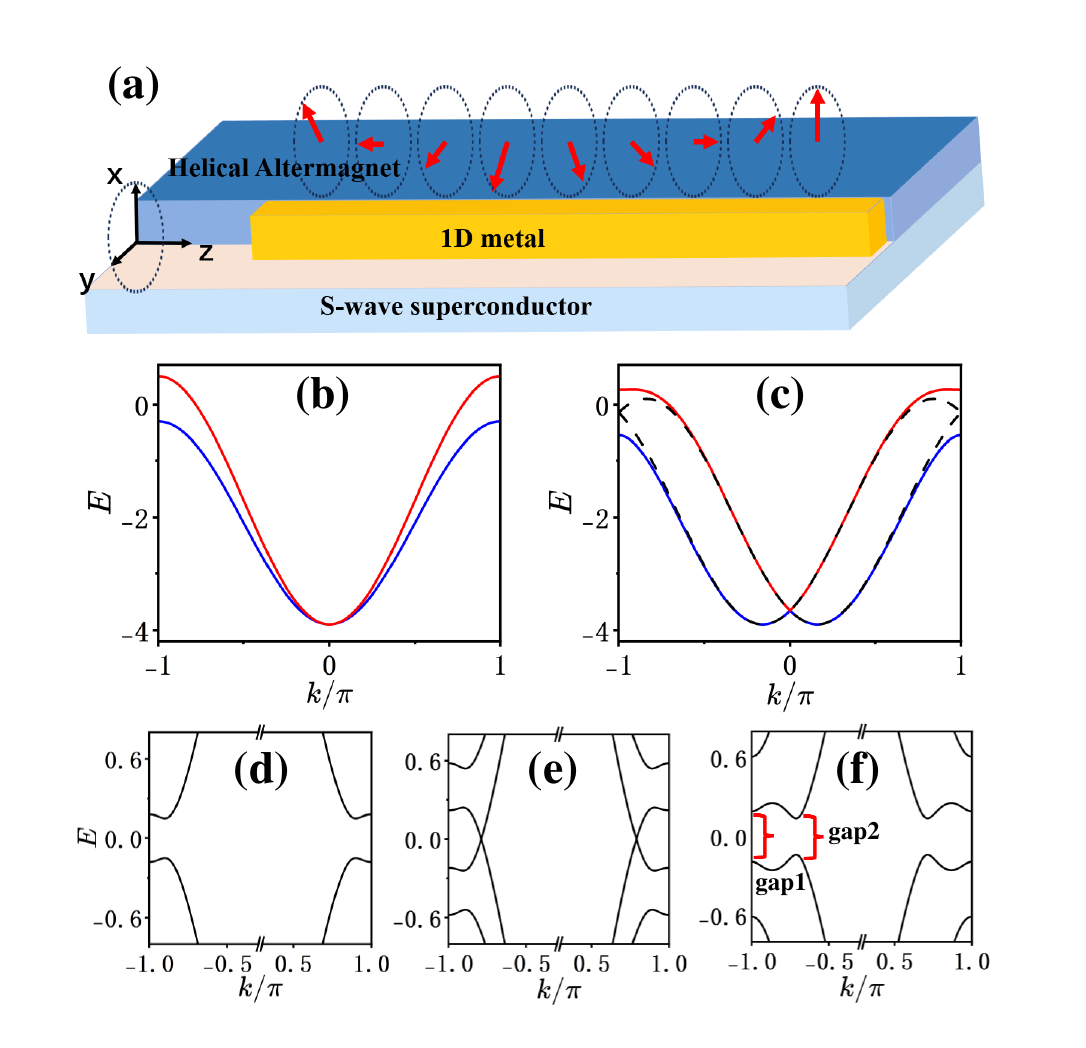}%
	\caption{\label{FIG1}
	(a) Schematic for altermagnet-metal-superconductor nanowire system to realize TSC, Néel vector of 2D altermagnet rotates in $xy$ plane versus $z$ coordinate.
	(b) Altermagnetic nanowire energy band without helical structure with parameters $\mu=1.9, \theta=0, t_J=0.1$.
	(c) The black dashed line is free electron energy band within rotating frame without altermagnetism. Parameters: $\mu=1.9, \theta=1, t_J=0$.
	The blue and red lines are energy bands of helical altermagnet nanowire with
	$t_J$ changed to $0.1$.
	(d), (e), and (f) are the energy bands of the nanowire in proximity to \emph{s}-wave superconductor without altermagnet and helical order ($\theta=0, t_J=0$),
with altermagnetism but without helical order ($\theta=0, t_J=0.1$),
and with both altermagnetism and helical order ($\theta=1, t_J=0.1$), respectively.
The parameters $\mu=1.9$ and $\Delta=0.15$.
	}
\end{figure}

In this Letter, we propose a scheme to realize MBSs
in a metal nanowire on the surface of a two-dimensional (2D) helical altermagnet and in proximity to an \emph{s}-wave SC without both SOC and net magnetism, as illustrated in Fig. \ref{FIG1}(a).
First, by gauge transformation and analysis of the band structure, we show that the helical altermagnetism hosts the spin splitting and opens a gap, possessing essential condition to support 1D TSC.
Next, we show that MBSs are localized at both ends of finite-size nanowire in proximity of an \emph{s}-wave SC and investigate the regulations of topological properties.
Furthermore, at the domain wall of nanowire where the helical chirality is inverted, we predict the existence of two MBSs protected by the chiral symmetry.

\emph{Model construction and energy band analysis}.-In low-energy (expanded near the $\Gamma$ point), the Hamiltonian of 1D metal nanowire along $z$ direction on the surface of 2D altermagnet can be written as $\frac{\hbar^2 k_z^2}{2m^*}-\tilde{\mu}+T_J k_z^2 \sigma_1$, where the last term generates spin splitting with altermagnet Néel vector along $x$ axis \cite{cheng_Fieldfree_2024, wei_Gapless_2024, yi_Spin_2025}. $k_z$ is the wave vector,  $m^*$ is effective electron mass, $\tilde{\mu}$ is the chemical potential, $T_J$ represents altermagnet strength and the Pauli matrices $\sigma_{1,2,3}$ acts on the spin space. Our model is shown in Fig. \ref{FIG1}(a), 1D metal nanowire along $z$ direction is located on the surface of 2D helical altermagnet and in proximity to an \emph{s}-wave SC. To characterize the helical altermagnetism, we substitute $\sigma_1$ with a location-dependent $\sigma_n(z) = \cos (\theta z) \sigma_1+ \sin (\theta z) \sigma_2$, which represents that the Néel vector rotates continuously in $xy$ plane and $\theta$ is the rotation frequency. 
In low-energy, the 1D metal nanowire in Fig. \ref{FIG1}(a) can be described by the Bogoliubov-de Gennes (BdG) Hamiltonian $H$ in the Nambu basis $\Psi=\{\psi_\uparrow, \psi_\downarrow, \psi_\uparrow^\dagger, \psi_\downarrow^\dagger\}^T$:
\begin{align}
	H=&\int   \Psi(z)^\dagger \left[ \left(-\frac{\hbar^2 \partial_z^2}{2m^*}-\tilde{\mu}\right)\sigma_0\tau_3-\frac{1}{2} T_J \left\{\partial_z^2,  \cos \left(\theta z \right) \right\}
	\sigma_1\tau_3  \nonumber \right.\\
	& \left.-\frac{1}{2} T_J \left\{ \partial_z^2, \sin \left(\theta z \right)\right\} \sigma_2\tau_0
	-\Delta\sigma_2\tau_2\right]  \Psi(z) dz, \label{E1}
\end{align}
where \{\} represents the anticommutator,
$\Delta$ represents proximity-induced
pairing potential, the Pauli matrices $\tau_{1,2,3}$ act on the particle-hole space.
In the numerical calculations, we use tight-binding lattice Hamiltonian: (For detailed derivation, see Sec. SI of the Supplementary Materials \cite{Sup}, see also references \cite{mao_Universal_2024, sun_Timedependent_1997, sun_Design_2023, sun_Design_2025, wingreen_Timedependent_1993, lee_Simple_1981, sancho_Highly_1985, sun_Quantum_2009, sun_Quantum_2011, jauho_Timedependent_1994, cheng_Controllable_2009, tewari_Topological_2012} therein.)
\begin{align}
	\mathcal{H}=&\sum_{j} \Psi_j^\dagger \left[-\mu\sigma_0\tau_3+2 t_J \left(\cos\left(\theta j \right) \sigma_1\tau_3+\sin(\theta j) \sigma_2\tau_0 \right)\nonumber  \right. \\&
	\left. -\Delta\sigma_2\tau_2\right]  \Psi_j +\Psi_j^\dagger\left[-t\sigma_0\tau_3-t_J \cos\left(\theta \left(j+1/2 \right)\right)\sigma_1\tau_3\nonumber \right.\\&\left.  -t_J\sin\left(\theta \left(j+1/2 \right)\right) \sigma_2\tau_0  \right]  \Psi_{j+1} +H.c.  . \label{E2}
\end{align}
Here, $\Psi_j=\{c_{j\uparrow}, c_{j\downarrow}, c_{j\uparrow}^\dagger, c_{j\downarrow}^\dagger\}^T$ in which $c_{j\alpha}^\dagger$ is the creation operator of electron with spin $\alpha$ on
lattice site $j$. $j=1,2,...N$ with $N$ the nanowire length.
Based on Eq. (\ref{E2}), we can derive its low-energy
effective Hamiltonian, given in Eq. (\ref{E1}). The corresponding parameters are related as $t=\frac{\hbar^2}{2m^{*} a^2}$, $t_J=\frac{T_J}{a^2}$, $\mu=\tilde{\mu}-2t$. Here we set the lattice constant $a=1$ as length unit.
Under this nondimensionalization, the angle between Néel vectors at adjacent sites becomes numerically equal to $\theta$. We set $t=1$ as energy unit in the following calculations.
Note that Eq. (\ref{E2}) is used for all the discrete calculations in this Letter. However, this Hamiltonian is still location-dependent, to determine the energy spectrum, we perform a location-dependent gauge transformation $c_{j\uparrow}\rightarrow c_{j\uparrow}e^{-i \theta j/2 }$ and $c_{j\downarrow} \rightarrow c_{j\downarrow}e^{i\theta j/2}$. In this new basis, the Hamiltonian in Eq. (\ref{E2}) can be rewritten as
\begin{align}
	\mathscr{H}&=\sum_{j} \Psi_j^\dagger\left(-\mu\sigma_0\tau_3+2t_J \sigma_1\tau_3-\Delta\sigma_2\tau_2\right)\Psi_j \nonumber\\&+ \Psi_j^\dagger (it \sin \frac{\theta}{2} \sigma_3 \tau_0
	-t \cos \frac{\theta}{2} \sigma_0 \tau_3
	-t_J \sigma_1 \tau_3 ) \Psi_{j+1} +H.c. . \label{E3}
\end{align}	

Let us give a physical picture of this transformation. The gauge transformation belongs to a local SU(2) unitary transformation, causing the coordinate axes of spin space to rotate in $xy$ plane. In such a rotating frame, the Néel vector of the helical altermagnet nanowire always remains fixed along the $x$-axis \cite{vazifeh_SelfOrganized_2013}.
In addition, considering the electron kinetic energy term which conserves electron's spin, in the rotating frame, the spin components in $xy$ plane now rotate, resulting in an intrinsic coupling between spin and momentum \cite{braunecker_Spinselective_2010, sun_Quantum_2005}, similar to SOC \cite{molenkamp_Rashba_2001}.

Under the spin local coordinate, the Hamiltonian in Eq. (\ref{E3}) has translational invariance now and can be Fourier transformed as
\begin{align}
	\mathscr{H}(k)&=
	\begin{bmatrix}
        H_1(k)&  i\sigma_2\Delta        \\
        -i\sigma_2\Delta &-H_1^*(-k)	
	\end{bmatrix} ,    \label{E4} \\
	H_1(k)&=
	\begin{bmatrix}
        -2t \cos (k-\theta/2)-\mu&  2t_J-2t_J \cos k        \\
        2t_J-2t_J \cos k &-2t \cos (k+\theta/2)-\mu	
	\end{bmatrix}.     \label{E5}
\end{align}
$H_1(k)$ is the helical altermagnet nanowire Hamiltonian without superconducting proximity effect \cite{fu_Superconducting_2008} and $\mathscr{H}(k)$ is the total Hamiltonian of 1D system.
According to $H_1(k)$, the nanowire energy band structure without SC is shown in Figs. \ref{FIG1}(b-c).
When $\theta=0$ (without helix), the spin degeneracy is lifted except $k=0$ and there is a spin splitting of $\pm4t_J$ at $k=\pi$ as shown in Fig. \ref{FIG1}(b).
When $\theta \neq0$, the band structure of helical altermagnet nanowire is demonstrated by color lines in Fig. \ref{FIG1}(c).
Through the Hamiltonian $H_1$, one can find the diagonal term is just the free electron's band structure which is translated by $\pm \theta/2$ from $-2t \cos(k)-\mu$, resulting in momentum being locked with $z$ component of spin. The non-diagonal term is just the altermagnet contribution in the periodic lattice whose Néel vector is along $x$ axis.
Now it is clear that the helical structure ($\theta$) provides the band splitting as demonstrated by the black dashed lines in Fig. \ref{FIG1}(c), and altermagnet ($t_J$) breaks the time-reversal symmetry \cite{smejkal_Emerging_2022, smejkal_Conventional_2022} and opens a gap of $\pm4t_J$ at $k=\pi$ as demonstrated by the color lines in Fig. \ref{FIG1}(c),
with the center energy of this gap is given by $2t \cos \theta/2-\mu$.
This 1D nanowire exhibits analogous characteristics to the semiconducting nanowire with SOC and magnetic field \cite{oreg_Helical_2010}.
%
%
Besides, according to the kitaev criterion \cite{kitaev_Unpaired_2001}, the system supports TSC when degeneracy is twofold around the Fermi level (zero energy) as shown in Fig. \ref{FIG1}(c). We can use this to control the topological phase. By tuning $\theta$ and $\mu$, we can adjust the relative position between the center of gap ($2t \cos \theta/2-\mu$) and Fermi level.
Simultaneously, $t_J$ can be regulated to control the size of the gap.
These factors collectively determine whether degeneracy around Fermi level is twofold, which is important for interpreting the phase diagram in subsequent sections. In contrast, in the semiconducting nanowire model, the strength of SOC cannot directly influence the phase transition \cite{klinovaja_Composite_2012, mishmash_Approaching_2016}.

Considering the proximity effect, according to $\mathscr{H}(k)$, the 1D system energy band structure is shown in Figs. \ref{FIG1}(d-f).
Figure \ref{FIG1}(d) illustrates energy band without helical altermagnetism, which exhibits spin degeneracy.
When $t_J$ increases from zero, the introduction of altermagnet term breaks the spin degeneracy and the SC gap decreases. When $t_J>t_J^c (t_J^c\approx\frac{\Delta}{2+\mu})$, the SC gap keeps closed as shown in Fig. \ref{FIG1}(e). (for detailed derivation, see Sec. SII of the Supplementary Materials \cite{Sup}.)
However, when $\theta \neq 0$ is introduced, the SC gap reopens, driving the system into TSC state, where MBSs are expected to emerge
as shown in Fig. \ref{FIG1}(f).
In this situation, two distinct gaps emerge. gap2 is mainly governed by the rotation frequency, gap1 (located at $k=\pi$) is associated with phase transition.
Different from former research \cite{lutchyn_Majorana_2010, oreg_Helical_2010}, the properties of altermagnet guarantee the phase transition only occurs at $k=\pi$ (because Kramers degeneracy is not lifted at $k=0$). According to band inversion at gap1 in Fig. \ref{FIG1}(f), TSC requires $\mu$ satisfies $(4t_J)^2>\Delta^2+(2t \cos \theta/2-\mu)^2$. (details shown in Sec. SII and Fig. S1 of the Supplementary Materials \cite{Sup}.)
This relation indicates that when Fermi level (zero energy) is positioned at center of the gap in Fig. \ref{FIG1}(c), $t_J$ required to support TSC is minimized.

If we take the lattice constant $a=6.6$ nm and the effective electron mass is set as $m^*= 0.08m_e$ with $m_e$ the bare electron mass ($\mathrm{Ge}$ and $\mathrm{GaAs}$ are candidate materials \cite{kittel_Introduction_1967}),
$t$ will be about 0.01 eV. Then $\Delta=0.15t$ corresponds to about $1.5$ meV and $t_J=0.1t$ corresponds to about $1$ meV, which are reasonable.

\begin{figure}
	\includegraphics[width=\columnwidth]{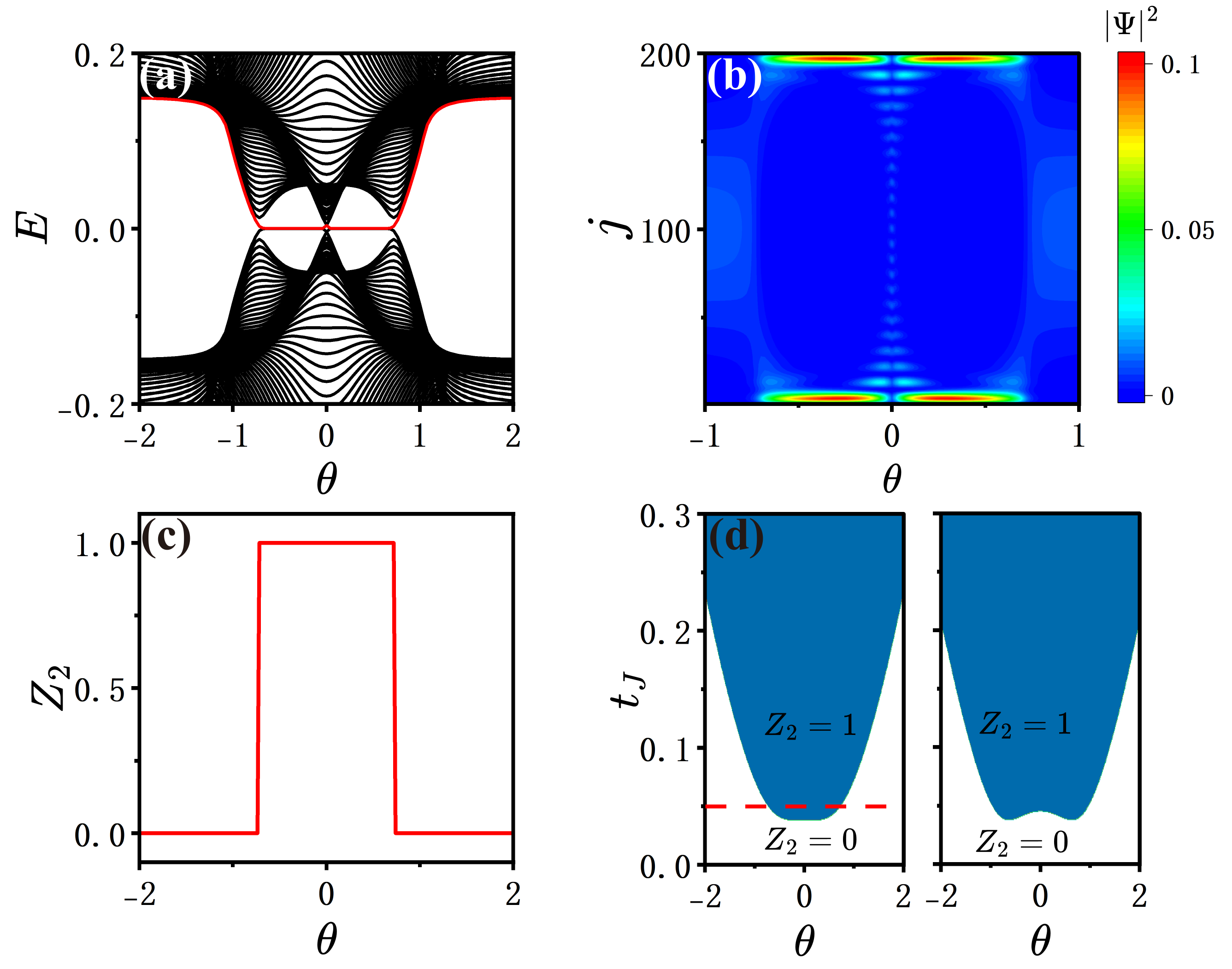}%
	\caption{\label{FIG2}
	(a) Finite-size energy spectrum of SC-proximitized helical altermagnet nanowire versus $\theta$.
	(b) Wave functions of the red lines in (a).
	(c) $Z_2$ number of the bulk Hamiltonian.
	(d) Topological phase diagram in the $t_J$-$\theta$ plane,
	$\mu=2$ and $1.9$ in the left and right panels.
	Parameters: $\Delta=0.15$ in (a-d), $\mu=2, t_J=0.05$ in (a-c), and the nanowire length $N=200$ in (a-b).}
\end{figure}

\emph{Topological properties in helical altermagnet nanowire}.-With the Hamiltonian of Eq. (\ref{E2}), Fig. \ref{FIG2}(a) demonstrates the eigenenergy of 1D finite-size helical altermagnet versus the rotation frequency $\theta$.
As $\theta$ increases from zero, the bulk energy increases from zero.
This behavior arises from the reopening and gradual widening of gap2 in Fig. \ref{FIG1}(f), analogous to the behavior dominated by SOC in semiconducting nanowire system \cite{klinovaja_Composite_2012, mishmash_Approaching_2016}.
Within the bulk gap, MBSs emerge [see the horizontal lines with zero energy Fig. \ref{FIG2}(a)].
With further increase of $\theta$, gap2 is larger than gap1, consequently, gap1 dominates the superconducting bulk gap.
Near the phase transition where $\theta \approx 0.75$ in Fig. \ref{FIG2}(a), band inversion of gap1 leads to the closing and reopening of the bulk energy gap, and then
the MBSs disappear.

In Fig. \ref{FIG2}(b), wave functions corresponding to red line in Fig. \ref{FIG2}(a) are plotted. It clearly shows that MBSs are localized at both ends of nanowire. When $\theta$ approaches zero, Majorana localization length increases longer than nanowire length, leading to the oscillatory behavior in the wave functions \cite{klinovaja_Composite_2012}. When $\theta=0$, gap2 is closed, there is no helical structure to protect TSC gap and induce MBSs.

We calculate $Z_2$ topological invariants versus $\theta$ using Pfaffian method \cite{fu_Time_2006} and Hamiltonian after gauge transformation [Eq. (\ref{E4})]
and show phase diagram in Figs. \ref{FIG2}(c,d).
The $Z_2=1$ regions all conform the result in Figs. \ref{FIG2}(a) and \ref{FIG2}(b) using the Hamiltonian before gauge transformation [Eq. (\ref{E2})].
The consistency of the two calculation results further validates
the existence of the MBSs.
However, at $\theta=0$, although $Z_2$ remains $1$, the gap closing and absence of MBSs in Figs. \ref{FIG2} (a) and \ref{FIG2}(b)
indicate a phase transition induced by the inversion of helical chirality.
This reveals the limitation of $Z_2$ classification and another topological classification is required.
We explain it in the following subsection.

In Fig. \ref{FIG2}(d), we present phase diagrams with different chemical potentials $\mu$. We focus on the phase boundary curve between $Z_2=1$ and $Z_2=0$.
The curves exhibit one minimum point in the left panel and the point splits into two in the right panel by tuning $\mu$. In fact, the horizontal positions of minimum points satisfy $2t \cos \theta/2-\mu=0$ in both panels.
This is because in Fig. \ref{FIG1}(c), when the Fermi level (zero energy) is located at the center of the gap ($2t \cos \theta/2-\mu$), time-reversal-breaking component ($t_J$) required to support TSC is minimal.
Along the boundary, when $\theta$ deviates from the minimum point $\theta_m$ ($\theta_m=2\arccos \frac{\mu}{2t}$ ), $t_J$ increase. This is because regulating $\theta$ shifts the gap in Fig. \ref{FIG1}(c) away from Fermi level, a larger $t_J$ is necessary to preserve the twofold degeneracy around the Fermi level.
Under the same parameters, the red dashed line is consistent with results in Figs. \ref{FIG2}(a-c).
Furthermore, both panels validate the topological criterion proposed earlier.

We also investigate the coexistence of helical structure and intrinsic SOC and find our mechanism remains valid (details shown in Sec. SIII and Fig. S2 of Supplementary Materials \cite{Sup}).
Besides, details on regulating $\mu$ and $t_J$ to support TSC are provided in Sec. SIV and Figs. S3-S4 of Supplementary Materials \cite{Sup}.

\begin{figure}
	\includegraphics[width=\columnwidth]{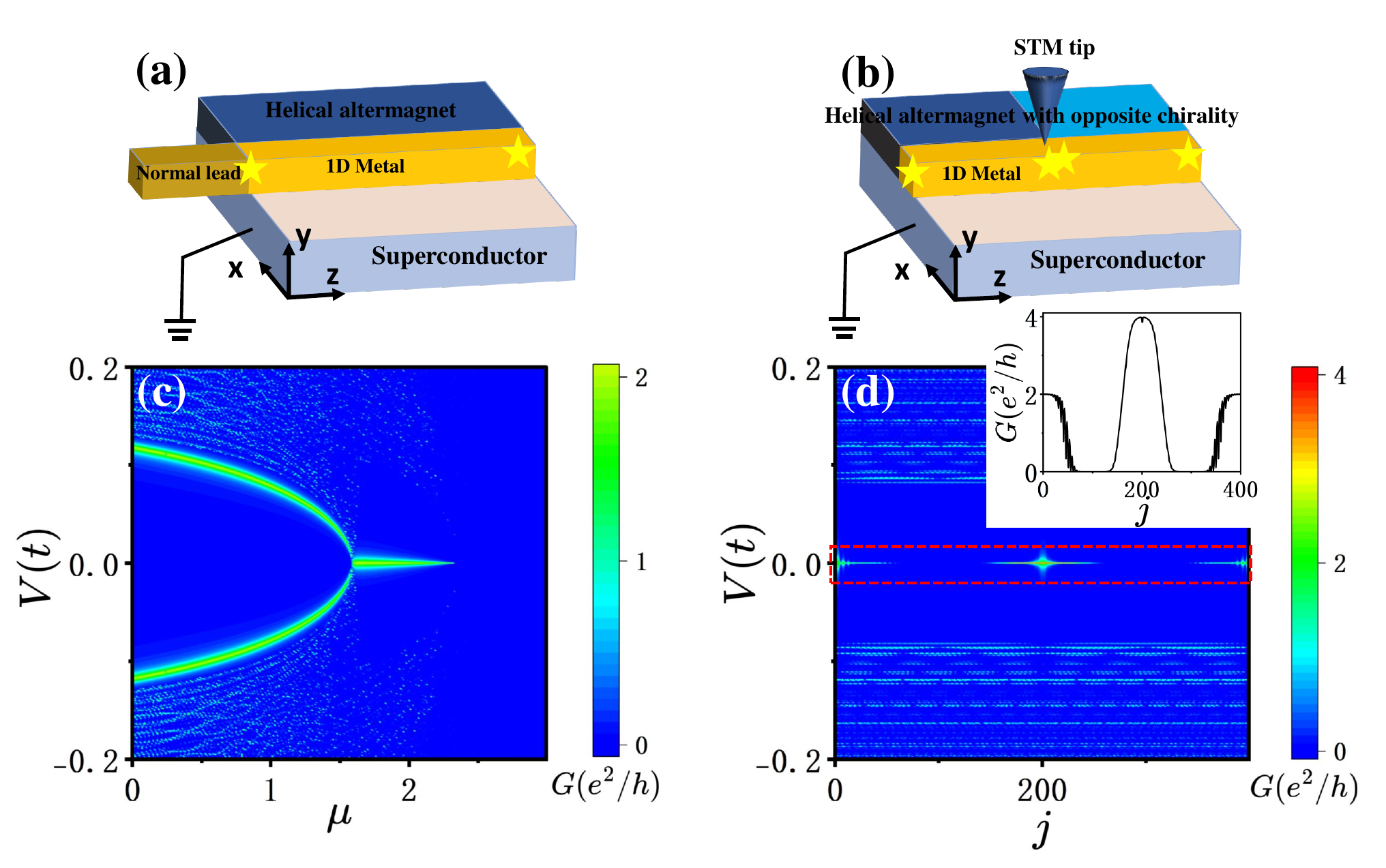}%
	\caption{\label{FIG3}
	(a) Two MBSs (yellow stars) are localized at both ends of nanowire with a metal lead connected to one end.
	(b) Helical altermagnet nanowire with opposite chirality. Two MBSs are localized at ends and the other two are localized at domain wall. The STM can scan the whole nanowire.
	(c) Calculated conductance for the device in (a) versus bias $V$ and $\mu$ with $N=200, \theta=0.4, t_J=0.1$.
	(d) Calculated STM conductance for the device in (b) versus bias $V$ and site $j$ with $t_J=0.1, \mu=2$. In the inset, conductance at zero bias is plotted.
The parameters $N=200, \theta=0.4$ for the left nanowire and $N=200, \theta=-0.4$
for the right nanowire.}
\end{figure}

\emph{Transport properties in helical altermagnet nanowire}.-As is well known,
one of the prominent feature of MBSs is the ZBCP through the resonant Andreev reflection \cite{law_Majorana_2009, flensberg_Tunneling_2010}.
We calculate the conductance of 1D helical altermagnet nanowire.
The schematic is shown in Fig. \ref{FIG3}(a) with two MBSs localized at ends of nanowire. A metal lead is used to contact and measure.
Using the Green's function method, conductance can be calculated as $G(V) = \frac{d\langle I_L \rangle}{dV}$, the detailed derivation is shown in Sec. SV of Supplementary Materials \cite{Sup}.

The conductance spectrum versus bias $V$ and chemical potential $\mu$ is shown in Fig. \ref{FIG3}(c). The range of $\mu$ where ZBCP emerges coincides the region of the existence of the MBSs [see Fig. S3(a) and Sec. SIV in the Supplementary Materials \cite{Sup}].
When $\mu$ lies in about $1.6-2.3$, ZBCP of a quantized value $2e^2/h$ is observed. When $\mu<1.6$, perfect ZBCP splits into two branches both exhibiting a large conductance, corresponding to two Andreev bound states (ABSs) in the SC gap without topological protection \cite{prada_Andreev_2020}. Wave functions of ABSs are also localized at ends of nanowire \cite{Sup}, providing a superconducting analog of obstructed atomic insulator \cite{khalaf_Boundaryobstructed_2021, liu_Massive_2024, xu_Fillingenforced_2024}. When $\mu>2.3$, the system is a trivial insulator with no conductance peaks.

As highlighted in Fig. \ref{FIG2}, the classification with $Z_2$ topological invariants fails to describe the topological phase transition from $\theta<0$ to $\theta>0$. A new classification is required to fully characterize topological properties \cite{ojanen_Topological_2013, poyhonen_Majorana_2014, schecter_Selforganized_2016, schnyder_Classification_2008}. Firstly, the Hamiltonian typically exhibits particle-hole symmetry $\{\mathscr{H},\Theta\}=0$, with $\Theta=\sigma_0 \tau_1 K$ where $K$ denotes complex conjugation. Meanwhile, in the case of coplanar helical altermagnet order, the Hamiltonian also possesses chiral symmetry $\{\mathscr{H}, C\}=0$ with $C=\sigma_1 \tau_2 $ and emergent time-reversal symmetry $[\mathscr{H}, T]=0$ with $T=\Theta C=-i\sigma_1\tau_3K$. Here, $[\,]$ represents the commutator. Thus, the system belongs to the BDI class in the topological classification \cite{schnyder_Classification_2008, ryu_Topological_2010}. Notably, this emergent time-reversal symmetry is distinct from the physical time-reversal symmetry with the operator: $T_{true}=i\sigma_2 K$, the latter is broken by the altermagnetic order in our model.
In Fig. \ref{FIG2}(c), the region where $Z_2=1$ can be further divided into two distinct regions depending on whether $\theta \in(-\pi,0)$ or $\theta \in(0,\pi)$, corresponding to different $Z$-valued topological invariants\cite{ojanen_Topological_2013, poyhonen_Majorana_2014}. We calculate the $Z$-valued topological invariants in Sec. SVI and Fig. S5 of Supplementary Materials \cite{Sup}.
So, when $\theta$ crosses zero, phase transition occurs accompanied by the closure of gap in Fig. \ref{FIG2}(a). Additionally, the domain wall between two TSCs where $\theta$ belongs to positive and negative topological regions respectively, is expected to host two MBSs.
These MBSs are protected by the chiral symmetry and can provide a $4e^2/h$ ZBCP through the resonant Andreev reflection \cite{law_Majorana_2009, flensberg_Tunneling_2010}.

To demonstrate this, we design a device and do a transport calculation. As shown in Fig. \ref{FIG3}(b), two helical altermagnet nanowires with opposite chirality are located on the surface of an \emph{s}-wave SC.
We simulate the scanning tunneling microscope (STM) as MBSs detector, which can be precisely positioned to measure the conductance of each site.
Detailed parameter settings are in Sec. SVII of Supplementary Materials \cite{Sup}.
The conductance spectrum versus bias $V$ and site $j$ is shown in Fig. \ref{FIG3}(d).
ZBCP emerges at the domain wall and on both sides.
As illustrated in the inset, the ZBCP on the sides exhibits conductance of $2e^2/h$, while the ZBCP at the domain wall shows a doubled value $4e^2/h$. This observation confirms the existence of two MBSs with opposite chirality that remain uncoupled.

We also investigate the impacts of disorder and find that transport properties exhibit robustness against disorder to some extent (details shown in Sec. SVIII and Fig. S6 of Supplementary Materials \cite{Sup}).

\emph{Discussion and Conclusion}.-Let us discuss the experimental prospects. The helical antiferromagnetic structure has been observed in various systems for two decades \cite{bode_Chiral_2007, yang_Spiraling_2000, wadley_Current_2018}.
For altermagnet, a recent experiment demonstrates the controlled formation of 100-nanometre-scale helical structure between different domains in $\mathrm{MnTe}$, a $g$-wave altermagnet \cite{amin_Nanoscale_2024}.
Thus, we also anticipate the existence of helical altermagnet in $d$-wave altermagnet.
Furthermore, numerous studies have investigated the manipulation of the helical antiferromagnet structure \cite{cheng_Dynamics_2014, gomonay_High_2016, tveten_Antiferromagnetic_2014}, such as controlling the motion of the structure by current \cite{baldrati_Mechanism_2019, janda_MagnetoSeebeck_2020, wadley_Current_2018} and laser dragging \cite{hedrich_Nanoscale_2021}.
For altermagnet, several experiments manipulate the alignment of the Néel vector in $\mathrm{RuO_2}$, a $d$-wave altermagnet \cite{bai_Observation_2022, zhang_Electrical_2025}. These findings provide valuable insights for manipulation of helical altermagnet order.

In conclusion, we propose that the 1D helical altermagnet nanowire is a good platform to realize TSC and MBSs.
Combined with band structure analysis, topological invariants, wave-function analysis and conductance calculations, we demonstrate the robust localization of MBSs at ends of the helical structure.
Notably, such a model simultaneously provides the two key ingredients required to support MBSs: time-reversal symmetry breaking and spin-momentum locking. More importantly, it requires neither spin-orbit coupling nor a net magnetic moment, thereby offering a more achievable and controllable platform for realizing MBSs.

\begin{acknowledgments}
  We thank Peng-Yi Liu, Yi-Xin Dai and Yu-Chen Zhuang for fruitful discussions.
	This work was financially supported by the National Key R and D Pro
	gram of China (Grant No. 2024YFA1409002), the National Natural Science Foundation of China (Grants No. 12374034 and No. 12447147), the 
	Quantum Science and Technology-National Science and Technology Major Project
(Grant No. 2021ZD0302403), and the China Postdoctoral Science Foundation (Grant No. 2024M760070). The computational resources are supported by High-performance Computing Platform of Peking University.
\end{acknowledgments}

\bibliography{refer.bib}

\end{document}


\renewcommand\thesection{S\Roman{section}}
\def\theequation{S\arabic{equation}}
\def\thefigure{S\arabic{figure}}

\title{Supplementary materials for ``Majorana modes in helical altermagnet without net magnetism and spin-orbit coupling"}

\author{Xing-Jian Yi}
\affiliation{International Center for Quantum Materials, School of Physics, Peking University, Beijing 100871, China}
\affiliation{Hefei National Laboratory, Hefei 230088, China}
\author{Yue Mao}
\affiliation{International Center for Quantum Materials, School of Physics, Peking University, Beijing 100871, China}
\author{Cheng-Ming Miao}
\affiliation{International Center for Quantum Materials, School of Physics, Peking University, Beijing 100871, China}
\author{Qing-Feng Sun}
\email[]{sunqf@pku.edu.cn}
\affiliation{International Center for Quantum Materials, School of Physics, Peking University, Beijing 100871, China}
\affiliation{Hefei National Laboratory, Hefei 230088, China}
\date{\today}

\maketitle
\tableofcontents
\section{\label{sec1}tight-binding lattice Hamiltonian}
In this section, we present the detailed derivation of Eq. (2) in the main text.
The Hamiltonian can be decomposed into kinetic energy term $H_{kinetic}$, altermagnet term $H_{helical}$ and superconducting pairing term $H_{\Delta}$. The tight-binding lattice Hamiltonians are given by \cite{ghorashi_Altermagnetic_2024}
\begin{align}
	H_{kinetic}=\sum_{j,s}  (-\mu)c_{j,s}^\dagger c_{j,s}-\sum_{j,s} (t c_{j,s}^\dagger c_{j+1,s}+H.c.), \label{S1}
\end{align}
\begin{align}
	H_{helical}&=\sum_{j,s,s'} 2t_J c_{j,s}^\dagger [\sigma_n(j)]_{s,s'}  c_{j,s'}  
	-\sum_{j,s,s'} t_J \left(c_{j,s}^\dagger [\sigma_n(j+1/2)]_{s,s'}   c_{j+1,s'} +H.c.   \right) , \label{S2}
\end{align}
\begin{align}
	H_{\Delta}=\sum_j (\Delta c_{j,\uparrow}^\dagger c_{j,\downarrow}^\dagger +H.c.) . \label{S3}
\end{align}
$c_{j,s}^\dagger$ ($c_{j,s}$)is the creation (annihilation) operator of electron with spin $s$ on lattice index $j$. $j=1,2,...N$.
$a$ is the lattice constant and $Na$ the nanowire length.
For the general altermagnet with Néel vector along $x$ direction, the Pauli matrices in Eq. (\ref{S2}) should be $\sigma_1$.
However, 
because of the helical order, the Néel vector orientation angle in the $xy$ plane is dependent on the location $z$ with $\sigma_n(z)=\cos (\theta z)\sigma_1+\sin (\theta z)\sigma_2$, where $\theta$ is helical frequency and $\theta z$ denotes the angle. In lattice, we replace continuum index $z$ by $ja$. Consequently, in Eq. (\ref{S2}), we use $\sigma_n(j)=\cos(\theta j)\sigma_1 +\sin(\theta j)\sigma_2$ for on-site term at site $j$
and $\sigma_n(j+1/2)=\cos(\theta (j+1/2))\sigma_1 +\sin(\theta (j+1/2))\sigma_2$ for hopping term between sites $j$ and $j+1$, where we have set $a=1$. By rewriting the Hamiltonian in Eqs. (\ref{S1}-\ref{S3}) into Nambu basis, we obtain the tight-binding Hamiltonian as shown in Eq. (2) in the main text.

\section{\label{sec3}Derivation of criterion}
Firstly, we present derivation of the critical value $t_J^c\approx\Delta/(2+\mu)$. As discussed in the second paragraph on page 3 in the main text, the SC gap keeps closed in the absence of helical structure when $t_J>t_J^c$.
Starting from Eq. (4) in the main text, we rewrite the Hamiltonian as
 \begin{align}
	\mathscr{H}(k)=
	 \begin{bmatrix}
  T+A &B&0&\Delta  \\
  B&T-A&-\Delta&0  \\
  0&-\Delta&-T+A&-B \\
  \Delta &0&-B&-T-A
\end{bmatrix}
	,       \label{S7}
\end{align}
where we have set $T=-t\cos (k-\theta/2)-t\cos (k+\theta/2)-\mu$, $A=-t\cos (k-\theta/2)+t\cos (k+\theta/2)$ and $B=2t_J-2t_J\cos k$.
By diagonalizing the Hamiltonian matrix, the energy eigenvalues are obtained
 \begin{align}
	E_{\pm}^2=T^2+A^2+B^2+\Delta^2\pm 2\sqrt{A^2T^2+B^2(\Delta^2+T^2)},
	      \label{S8}
\end{align}
where $E_{+}$ denotes two bands far from zero energy and $E_{-}$ denotes two bands near zero energy.
In the absence of helical structure ($\theta=0$), we have $A=0$. By setting $E_{-}=0$ in Eq. (\ref{S8}), we obtain the relation $B^2=\Delta^2+T^2$.
By substituting $T=-2\cos k-\mu$ $(t=1)$ and $B=2t_J-2t_J\cos k$,
we obtain following quadratic equation in terms of $\cos k$
 \begin{align}
	(4t_J^2-4)\cos k^2-(8t_J^2+4\mu)\cos k+ (4t_J^2-\Delta^2-\mu^2)=0
	.       \label{S9}
\end{align}
This equation may have $0$, 2, or 4 real solutions within the first Brillouin zone $k\in[-\pi,\pi]$. At the instant when gap closes, the equation yields two real solutions. By setting the discriminant of the quadratic equation to zero, the critical condition is derived as ${t_J^c}^2[(\mu+2)^2+\Delta^2]=\Delta^2$,
from which we obtain the critical value ${t_J^c}^2=\Delta^2/[(\mu+2)^2+\Delta^2]$.
Given that $\Delta^2 \ll (\mu+2)^2$, the expression simplifies to $t_J^c\approx\Delta/(2+\mu)$.

\begin{figure}
	\includegraphics[width=\columnwidth]{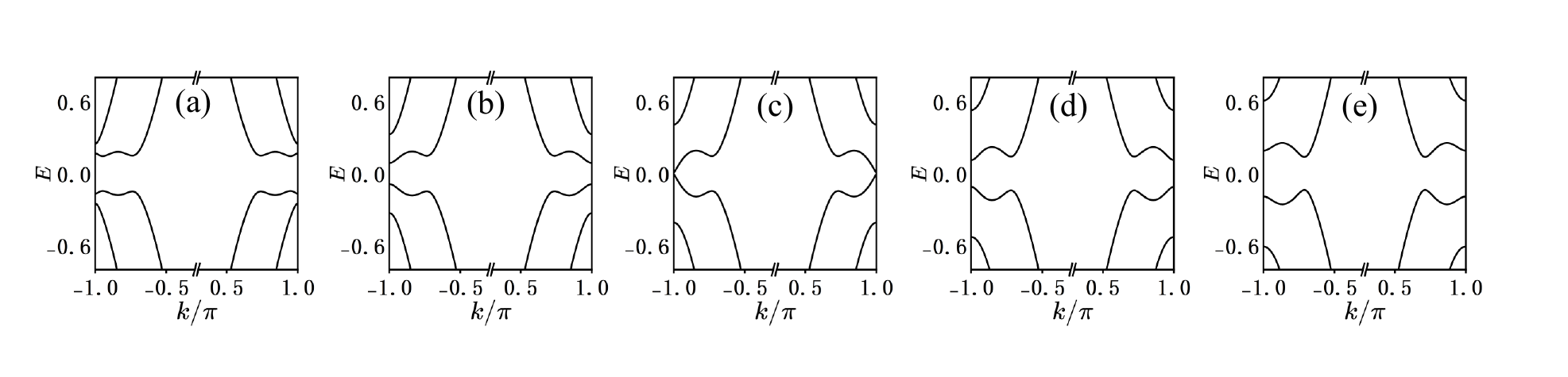}
	\caption{\label{FIGS1}
	\baselineskip 14pt
\sl{   (a-e) Band inversion process. 
Parameters: $t_J=0.01$ in (a), $t_J=0.03$ in (b), $t_J=0.05$ in (c), $t_J=0.08$ in (d), $t_J=0.1$ in (e), and the other parameters are the same as those in Fig. 1(f) in the main text.
}}
\end{figure}

Secondly, we present the derivation of phase transition criterion.
Figure \ref{FIGS1} demonstrates the band inversion process, manifested by the closure and subsequent reopening of the energy gap at the high-symmetry point $k=\pi$ (``gap1" in the main text) when the altermagnet strength $t_J$ increases.
At $k=\pi$, $A$ vanishes ($A=0$).
Utilizing the gap-closing condition ($E_{-}(k=\pi)=0$), we obtain $B^2=\Delta^2+T^2$, derived from Eq. (\ref{S8}).
By substituting $T=2t\cos \theta/2-\mu$ and $B=4t_J$,
we obtain the phase transition criterion $(4t_J)^2=\Delta^2+(2t\cos\theta/2-\mu)^2$.

\section{\label{sec4}Topological properties in helical altermagnet with SOC} 
In the main text, we propose that helical structure can induce the spin-momentum locking and effectively eliminate the requirement of SOC.
Our central findings remain valid even in the presence of SOC.
In the following, we systematically investigate how the coexistence of helical structure and intrinsic SOC affects our findings. We consider the SOC with the form $k_z\sigma_3 \tau_0$.

In the presence of SOC, the total Hamiltonian within the tight-binding model is given by
 \begin{align}
	H_{TOTAL}=H_{SOC}+\mathcal{H}
	,       \label{S100}
\end{align}
where
\begin{align}
	\mathcal{H}=&\sum_{j} \Psi_j^\dagger \left[-\mu\sigma_0\tau_3+2 t_J \left(\cos\left(\theta j \right) \sigma_1\tau_3+\sin(\theta j) \sigma_2\tau_0 \right)\nonumber  \right. \\& \left. -\Delta\sigma_2\tau_2\right]  \Psi_j +\bigg[\Psi_j^\dagger\left[-t\sigma_0\tau_3-t_J \cos\left(\theta \left(j+1/2 \right)\right)\sigma_1\tau_3\nonumber \right.\\&\left.  -t_J\sin\left(\theta \left(j+1/2 \right)\right) \sigma_2\tau_0  \right]  \Psi_{j+1} +H.c.\bigg]   , \label{S101}
\end{align}
and
 \begin{align}
	H_{SOC}=\sum_{j} \Psi_j^\dagger \left[(-i\alpha/{2a})\sigma_3\tau_0 \right]\Psi_{j+1} +H.c.
	       \label{S102}
\end{align}
Here, $\Psi_j=\{c_{j\uparrow}, c_{j\downarrow}, c_{j\uparrow}^\dagger, c_{j\downarrow}^\dagger\}^T$ in which $c_{js}^\dagger$ is the creation operator of electron with spin $s$ on
lattice site $j$ and $a=1$ is set as length unit.
$\mathcal{H}$ is the nanowire Hamiltonian, identical to Eq. (2) in the main text. $H_{SOC}$ is the SOC term.
By the location-dependent gauge transformation in the main text, $c_{j\uparrow}\rightarrow c_{j\uparrow} e^{-i\theta j/2}$ and $c_{j\downarrow}\rightarrow c_{j\downarrow} e^{i\theta j/2}$,
$\mathcal{H}$ is transformed as shown in the main text,
and $H_{SOC}$ can be reduced to
\begin{align}
	&\mathcal{H_{SOC}}=\sum_{j} \Psi_j^\dagger \hat{T}\Psi_{j+1} +H.c.   \nonumber \\
	\hat{T}&=diag\left(-i\frac{\alpha}{2}e^{-i\frac{\theta}{2}}, i\frac{\alpha}{2}e^{i\frac{\theta}{2}}, -i\frac{\alpha}{2}e^{i\frac{\theta}{2}} , i\frac{\alpha}{2}e^{-i\frac{\theta}{2}}   \right)
	        \label{S11}
\end{align}
Then $\mathcal{H_{SOC}}$ can be Fourier
transformed as
\begin{align}
	\mathcal{H_{SOC}}(k)=
	 diag \left(\alpha \sin(k-\frac{\theta}{2}), -\alpha\sin(k+\frac{\theta}{2}), \alpha\sin(k+\frac{\theta}{2}), -\alpha\sin(k-\frac{\theta}{2}) \right) 
	.        \label{S12}
\end{align}
Along with Eq. (4) and Eq. (5) in the main text, the total Hamiltonian including SOC is given by
\begin{align}
	\mathscr{H}_{TOTAL}(k)&=
	\begin{bmatrix}
        H_2(k)&  i\sigma_2\Delta        \\
        -i\sigma_2\Delta &-H_2^*(-k)	
	\end{bmatrix} ,    \label{S103} 
\end{align}
\begin{align}
	H_2(k)&=
	 \begin{bmatrix}
  -2\cos(k-\frac{\theta}{2})-\mu+ \alpha \sin(k-\frac{\theta}{2}) &2t_J-2t_J\cos k  \\
  2t_J-2t_J\cos k  &  -2\cos(k+\frac{\theta}{2})-\mu-\alpha \sin(k+\frac{\theta}{2})  
\end{bmatrix}   \nonumber \\
	&= \begin{bmatrix}
  -\sqrt{4+\alpha^2} \cos(k-\frac{\theta}{2}+\arctan \frac{\alpha}{2})-\mu &2t_J-2t_J\cos k  \\
  2t_J-2t_J\cos k  &  -\sqrt{4+\alpha^2} \cos(k+\frac{\theta}{2}-\arctan \frac{\alpha}{2})-\mu  
\end{bmatrix}     ,        \label{S13}
\end{align}
where we have set $t=1$ as energy unit. From Eq. (\ref{S13}), one can find 
the diagonal term is just the free electron's band structure which is translated by ($\pm\theta/2 \mp \arctan \frac{\alpha}{2}$) from $-\sqrt{4+\alpha^2}\cos k-\mu$, resulting in momentum being locked with $z$ component of spin.
The SOC is equivalent to a modification of the helical order, replacing $\theta/2$ by $\theta/2-\arctan (\alpha/2)$.
When $\arctan (\alpha/2)=\theta/2$, there is no spin-momentum locking to protect TSC gap and induce
MBSs.

\begin{figure}
	\includegraphics[width=\columnwidth]{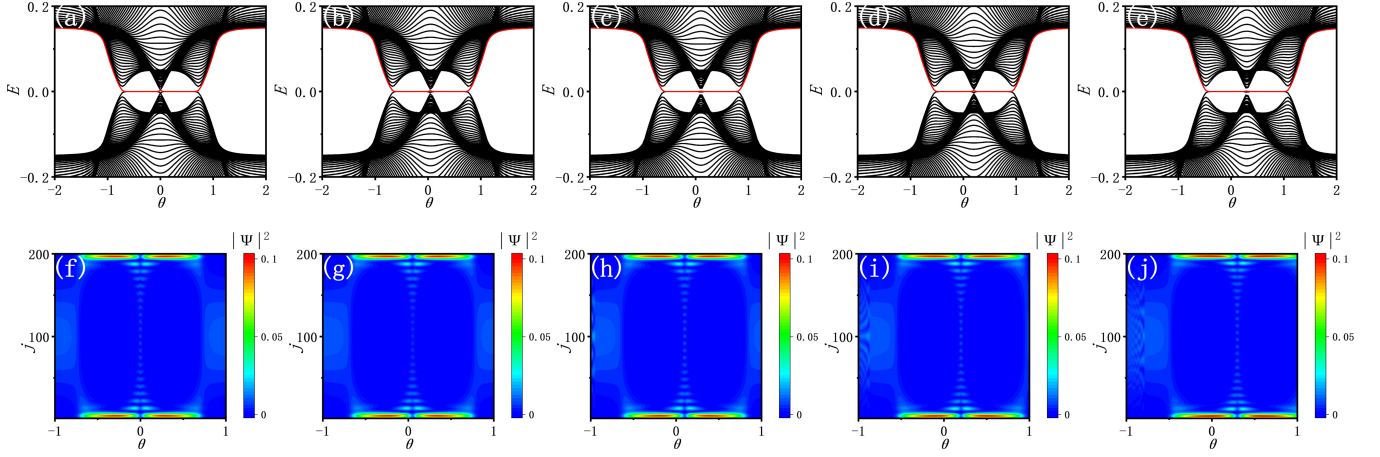}
	\caption{\label{FIGS2}
	\baselineskip 14pt
\sl{   (a-e) Finite-size energy spectrum of SC-proximitized helical altermagnet nanowire versus $\theta$. (f-j) Wave functions of
the red lines in (a-e). 
Parameters: $\alpha=0$ in (a,f), $\alpha=0.05$ in (b,g), $\alpha=0.1$ in (c,h), $\alpha=0.2$ in (d,i), $\alpha=0.3$ in (e,j), and the other parameters are the same as those in Figs. 2(a-b) in the main text.
}}
\end{figure}

To verify our derivation, we calculate the energy spectrum and wave functions in Fig. \ref{FIGS2} using the Hamiltonian Eq. (\ref{S100}).
Here, $\alpha=0.3$ corresponds to about $0.2$ eV $\mathrm{\mathring{A}}$ \cite{mourik_Signatures_2012}, a high SOC strength. Thus, we choose five $\alpha$ values from 0 to 0.3.
Overall, SOC shifts the energy spectrum, leading to the emergence of MBSs even in the absence of helical structure ($\theta=0$), consistent with previous studies \cite{oreg_Helical_2010, lutchyn_Majorana_2010}.
When $\theta$ and $\alpha$ have opposite signs, their effects superimpose and collectively enhance the spin-momentum locking.
Besides, the energy gap closes at $\theta \approx\alpha$ for all values of $\alpha$ in Figs. \ref{FIGS2}(a-e) and MBSs disappear at ends of the nanowire in Figs. \ref{FIGS2}(f-j), which are consistent with our analysis above.

In summary, the helical structure alone is sufficient to induce MBSs, eliminating the requirement for SOC. Moreover, even in the presence of SOC, our central conclusions are still valid: the helical structure can enhance spin-momentum locking and induce MBSs at ends of nanowire.

\section{\label{sec5}Other study of topological properties of TSC}
\begin{figure}
	\includegraphics[width=\columnwidth]{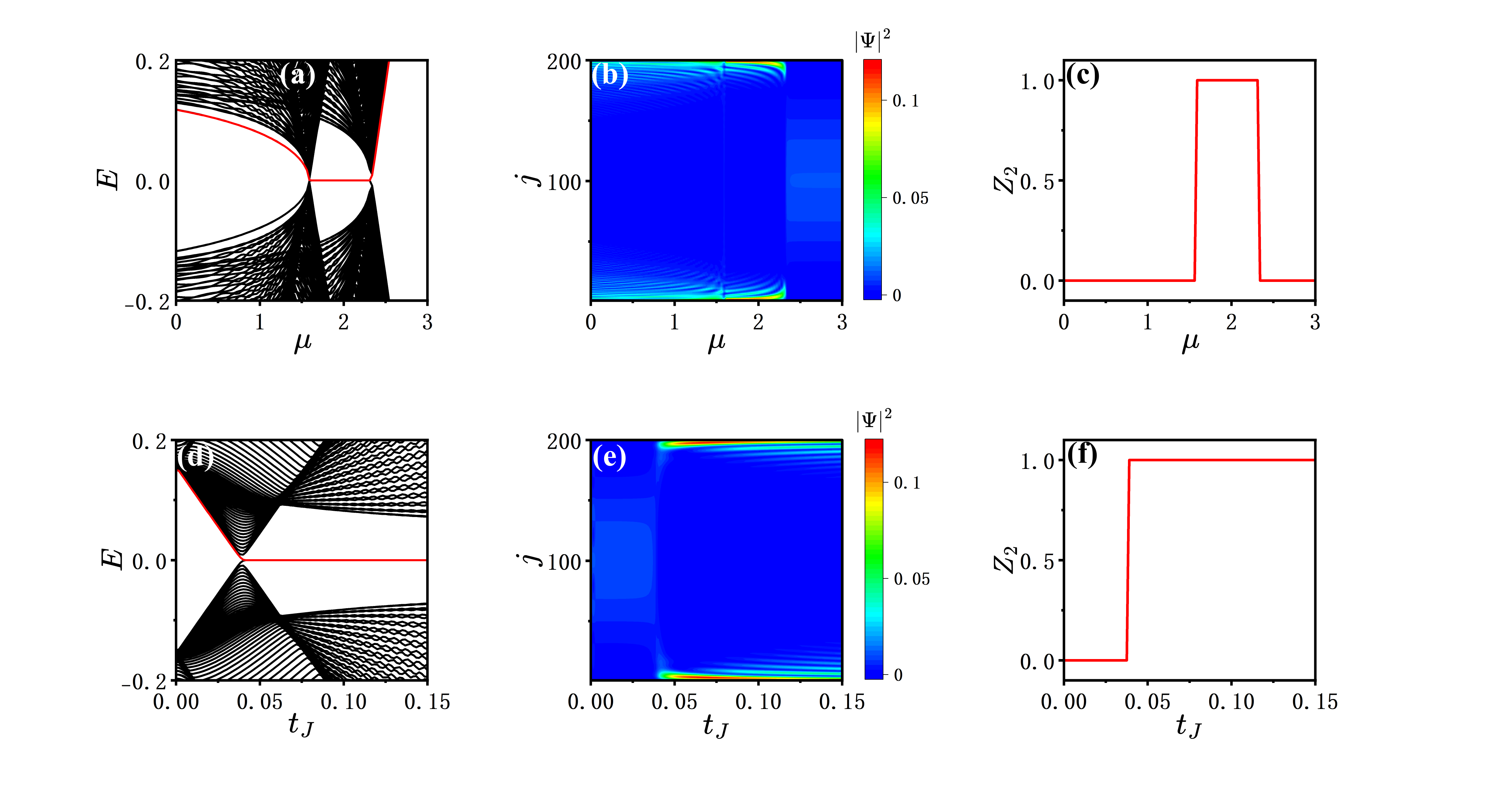}
	\caption{\label{FIGS3}
	\baselineskip 14pt
\sl{ (a) Finite-size energy spectrum of SC-proximitized helical altermagnet nanowire versus $\mu$ with the nanowire length $N=200$.
(b) Wave functions of the red line in panel (a).
(c) $Z_2$ topological invariants of the bulk Hamiltonian versus $\mu$.
(d) Finite-size energy spectrum of SC-proximitized helical altermagnet nanowire versus $t_J$ with the nanowire length $N=200$.
(e) Wave functions of the red line in panel (d).
(f) $Z_2$ topological invariants of the bulk Hamiltonian versus $t_J$.
Parameters: $\theta=0.4$, $\Delta=0.15$, $t_J=0.1$ in (a-c) and $\theta=0.4$, $\Delta=0.15$, $\mu=2$ in (d-f).}}
\end{figure}

Figure \ref{FIGS3}(a) demonstrates the eigenenergy of the finite-size 1D TSC versus chemical potential $\mu$ and MBSs are founded when $\mu$ lies in about $1.6-2.3$.  Meanwhile, its wave functions are localized at two ends of the nanowire as shown in Fig. \ref{FIGS3}(b). The $Z_2$ topological invariant is shown in Fig. \ref{FIGS3}(c), where parameter range with $Z_2=1$ is consistent with results in Fig. \ref{FIGS3}(a) and \ref{FIGS3}(b), indicating that the system is in a TSC phase. The above results are all consistent with the parameter range where the ZBCP appears in Fig. 3(c) in the main text.
Fig. \ref{FIGS3}(d) demonstrates the eigenenergy of the finite-size 1D TSC versus altermagnet strength $t_J$. The wave function of MBSs is also shown in Fig. \ref{FIGS3}(e). The $Z_2$ topological invariant shown in Fig. \ref{FIGS3}(f) is also consistent with results in Fig. \ref{FIGS3}(d) and \ref{FIGS3}(e).
It worth noting that in Fig. \ref{FIGS3}(a), when $\mu$ is less than about 1.6, there are two Andreev bound states in the superconducting gap without topological protection. The wave function is also localized at two ends of the nanowire which provides a superconducting analog of obstructed atomic insulator \cite{khalaf_Boundaryobstructed_2021, liu_Massive_2024, xu_Fillingenforced_2024}.

\begin{figure}[htbp]
	\includegraphics[width=0.6\columnwidth]{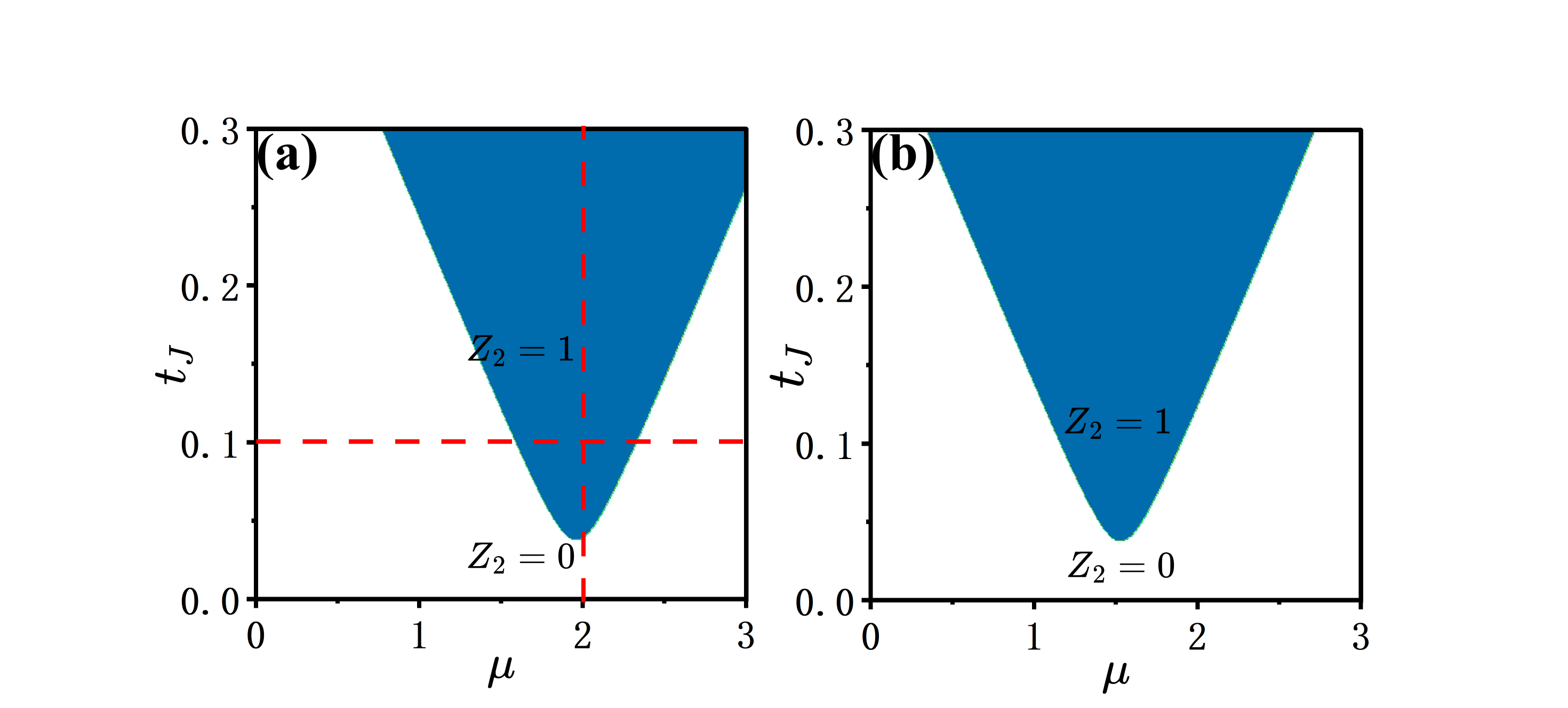}%
	\caption{\label{FIGS4}
	\baselineskip 14pt
	{\sl
	Topological phase diagram in the $t_J-\mu$ plane. Parameters: $\theta=0.4$ in (a), $\theta=1.4$ in (b), and $\Delta=0.15$ in (a) and (b).}}
\end{figure}

In Figs. \ref{FIGS4}(a) and \ref{FIGS4}(b), we plot phase diagram with different $\theta$ in the $t_J$ and $\mu$ plane. We focus on the phase boundary curve between $Z_2=1$ and $Z_2=0$.
At the minimum point of curve where $t_J$ is minimal, the horizontal position all satisfy the relation $2t\cos \theta-\mu=0$. This is because when Fermi level (zero energy) is in the center of the gap in Fig. 1(c) of the main text, time-breaking component ($t_J$) needed to enter TSC is the lowest. Along the boundary, $t_J$ increases when $\mu$ is far from the minimum point. That is because the center of the gap ($2t\cos \theta-\mu$) gradually moves away from Fermi level by regulating $\mu$, so a larger $t_J$ is required to guarantee the degeneracy around Fermi level to be twofold.
Under the same parameters,
the horizontal and vertical red dotted lines in Fig. \ref{FIGS4}(a) align with the results in Figs. \ref{FIGS3}(a-c) and Figs. \ref{FIGS3}(d-f).

\section{\label{sec6}Derive of current and conductance}
About our transport model in Fig. 3(a) in the main text, a metal lead is linked to the helical altermagnet nanowire with weak coupling.
The total Hamiltonian writes
\begin{eqnarray}
	H_{total}=H_{L}+H_{nw}+H_T. \label{S14}
\end{eqnarray}
where the $H_{L}$, $H_{nw}$ and $H_T$ represent the lead Hamiltonian, 1D helical altermagnet nanowire Hamiltonian and tunneling Hamiltonian between lead and 1D nanowire.
The lead Hamiltonian writes
\begin{eqnarray}
	H_{L}= \frac{\hbar^2 k^2}{2m^*}-\mu_0   , \label{S15}
\end{eqnarray}
where $m^*$ is the effective electron mass, $\mu_0$ is the lead's chemical potential.
In the Nambu basis $c_j=\{c_{j\uparrow}, c_{j\downarrow}, c_{j\uparrow}^\dagger, c_{j\uparrow}^\dagger\}^T$ with $c_{js}^\dagger$ ($c_{js}$) the creation (annihilation) operator at site $j$ with spin $s$, we discretize it as
\begin{eqnarray}
	H_{L}=\sum _{j=0}^{-\infty} c_j^\dagger (t_l \sigma_0 \tau_3)c_j +\sum _{j=-1}^{-\infty} (c_j^\dagger (-t_l \sigma_0 \tau_3) c_{j+1}+h.c.)   , \label{S16}
\end{eqnarray}
where we have set $t_l=\frac{\hbar^2}{2m^*a^2}$, $\mu_0=t_l$. When calculating, we set $t_l=1$.
We also set the lattice index of the lead as $j\leq0$ and that of 1D nanowire as $j>0$. About the 1D nanowire Hamiltonian $H_{nw}$, we have given in Eq. (2) in the main text.
About the tunneling Hamiltonian
\begin{eqnarray}
	H_T=c_0^\dagger T_{01} c_1+c_1^\dagger T_{10} c_0  , \label{S17}
\end{eqnarray}
with $	T_{01}=-t_M \sigma_0 \tau_3$ and $T_{10}=T_{01}^\dagger$.
We consider weak coupling and set $t_M=0.2$.
The current from the normal lead to the central region can be calculated from the evolution of the number operator of the electrons in the normal lead \cite{mao_Universal_2024, sun_Timedependent_1997}.
\begin{align}
    I_e&=-e \big<\frac{d(N_{lead\uparrow}+N_{lead\downarrow})}{dt} \big>=\frac{ie}{\hbar}\big<[\sum_{n=0}^{{-\infty}}(c_{n\uparrow}^\dagger c_{n\uparrow}+c_{n\downarrow}^\dagger c_{n\downarrow}),\sum_\sigma (t_M c_{0\sigma}^\dagger c_{1\sigma}+t_M c_{1\sigma}^\dagger c_{0\sigma})] \big>  \nonumber \\ 
	&=\frac{ie}{\hbar} \sum_\sigma (t_M\langle c_{0\sigma}^\dagger c_{1\sigma}\rangle - t_M\langle c_{1\sigma}^\dagger c_{0\sigma}\rangle)\nonumber   \\  
	&=\frac{e}{\hbar}  \mathrm{Tr}\{\Gamma_c [ \mathbf{G}_{10}^<(t,t) T_{01}-T_{10} \mathbf{G}_{01}^<(t,t)]\} \nonumber\\
    &=\frac{e}{h} \int d\omega \mathrm{Tr}\{\Gamma_c [ \mathbf{G}_{10}^<(\omega)T_{01}-T_{10} \mathbf{G}_{01}^<(\omega)]\} ,\label{S18}
\end{align}
where $\Gamma_c=diag(1,1,0,0)$.

The Green's functions are defined in the four component Nambu basis as \cite{sun_Design_2023, sun_Design_2025}
\begin{align}
	\mathbf{G}_{i,j}^r(t,t')=-i \theta(t-t')  \left< \left \{
    \begin{pmatrix}
        c_{i\uparrow}(t) \\
        c_{i\downarrow}(t) \\
        c_{i\uparrow}^\dagger(t) \\
        c_{i\downarrow}^\dagger(t)
    \end{pmatrix}
    \begin{pmatrix}
            c_{j\uparrow}^\dagger(t') &c_{j\downarrow}^\dagger(t') &c_{j\uparrow}(t') &c_{j\downarrow}(t')
    \end{pmatrix} \right  \} \right>  ,  \nonumber \\
    \mathbf{G}_{i,j}^<(t,t')=i
        \begin{pmatrix}
            \langle c_{j\uparrow}^\dagger(t') c_{i\uparrow}(t)\rangle & \langle c_{j\downarrow}^\dagger(t') c_{i\uparrow}(t) \rangle & \langle c_{j\uparrow}(t') c_{i\uparrow}(t)\rangle & \langle c_{j\downarrow}(t') c_{i\uparrow}(t)\rangle\\
            \langle c_{j\uparrow}^\dagger(t') c_{i\downarrow}(t)\rangle & \langle c_{j\downarrow}^\dagger(t') c_{i\downarrow}(t) \rangle & \langle c_{j\uparrow}(t') c_{i\downarrow}(t) \rangle & \langle c_{j\downarrow}(t') c_{i\downarrow}(t)\rangle \\
            \langle c_{j\uparrow}^\dagger(t') c_{i\uparrow}^\dagger(t)\rangle & \langle c_{j\downarrow}^\dagger(t') c_{i\uparrow}^\dagger(t)\rangle & \langle c_{j\uparrow}(t') c_{i\uparrow}^\dagger(t)\rangle & \langle c_{j\downarrow}(t') c_{i\uparrow}^\dagger(t)\rangle \\
            \langle c_{j\uparrow}^\dagger(t') c_{i\downarrow}^\dagger(t)\rangle & \langle c_{j\downarrow}^\dagger(t') c_{i\downarrow}^\dagger(t)\rangle & \langle c_{j\uparrow}(t') c_{i\downarrow}^\dagger(t)\rangle & \langle c_{j\downarrow}(t') c_{i\downarrow}^\dagger(t)\rangle
        \end{pmatrix}.  \label{S19}
\end{align}
$\mathbf{G}_{10}^<(\omega)$ and $\mathbf{G}_{01}^<(\omega)$ are obtained by using the Dyson equation \cite{wingreen_Timedependent_1993, mao_Universal_2024, sun_Timedependent_1997},
$\mathbf{G}_{10}^<=\mathbf{G}_{11}^r \bm{\Sigma}_{10}^r \mathbf{g}_{00}^<+\mathbf{G}_{11}^< \bm{\Sigma}_{10}^a \mathbf{g}_{00}^a ,
\mathbf{G}_{01}^<=\mathbf{g}_{00}^r \bm{\Sigma}_{01}^r \mathbf{G}_{11}^<+\mathbf{g}_{00}^< \bm{\Sigma}_{01}^a \mathbf{G}_{11}^a $, where $\bm{\Sigma}_{01}^r=T_{01}$, $\bm{\Sigma}_{10}^r=T_{10}$, $\bm{\Sigma}_{01}^a=[\bm{\Sigma}_{10}^r]^\dagger=T_{10}^\dagger=T_{01}$, $\bm{\Sigma}_{10}^a=[\bm{\Sigma}_{01}^r]^\dagger=T_{01}^\dagger=T_{10}$, $g_{00}^r$ is the surface Green's function of the lead \cite{lee_Simple_1981, sancho_Highly_1985}.
Then substitute the relations in Eq. (\ref{S18}) \cite{sun_Quantum_2009, sun_Quantum_2011},
\begin{align}
    I_e
    &=\frac{e}{h} \int d\omega \mathrm{Tr}\{\Gamma_c [(\bm{\Sigma}_0^a-\bm{\Sigma}_0^r)\mathbf{G}_{11}^<+\bm{\Sigma}_0^<(\mathbf{G}_{11}^r-\mathbf{G}_{11}^a)  ]\}     ,\label{S20}
\end{align}
where ${T}_{10} \mathbf{g}_{00}^< T_{01}=\bm{\Sigma}_0^<$,
$T_{10} \mathbf{g}_{00}^a T_{01}=\bm{\Sigma}_0^a$,
and $T_{10} \mathbf{g}_{00}^r T_{01}=\bm{\Sigma}_0^r$.
$\bm{\Sigma}^{r,a,<}_0 (\omega)$ are the retarded, advanced and lesser self-energies of coupling to the lead.
They are all diagonal in Nambu space.
Here $\bm{\Sigma}_{0}^<=- \mathbf{f}(\bm{\Sigma}_{0}^r-\bm{\Sigma}_{0}^a)= i  \mathbf{f} \bm{\Gamma}_{0}$ and $\bm{\Gamma}_0$ is defined as $\bm{\Gamma}_{0}=i(\bm{\Sigma}_{0}^r-\bm{\Sigma}_{0}^a)$. The Fermi distribution matrix $\mathbf{f}$ for the normal lead can be witten as
\begin{align}
	\mathbf{f}&=
    \begin{pmatrix}
		f_e(\omega) &0&0&0 \\
		0 & f_e(\omega) &0&0  \\
		0&0& f_h(\omega)&0  \\
		0&0&0 & f_h(\omega)
	\end{pmatrix}   ,\label{S21}
\end{align}
where $f_e(\omega)=f(\omega-eV)$, $f_h(\omega)=f(\omega+eV)$ with $V$ the bias voltage.
$f(\omega)=1/\{exp[(\omega-E_F)/{k_BT}] +1\}$ is the Fermi distribution function, where  $T$ is the temperature and Fermi level $E_F$ is set to be zero to maintain consistency with the main text. 
Then the charge current can be reduced to
\begin{align}
    I_c&=\frac{ie}{h} \int d\omega \mathrm{Tr}\Gamma_c[\bm{\Gamma}_{0} \mathbf{G}_{11}^<+ \mathbf{f} \bm{\Gamma}_{0} (\mathbf{G}_{11}^r-\mathbf{G}_{11}^a) ]   ,\label{S22}
\end{align}
Then by using the Keldysh equation \cite{jauho_Timedependent_1994}, we have $\mathbf{G}_{11}^<=\mathbf{G}_{11}^r \bm{\Sigma}_0^<\mathbf{G}_{11}^a$,
$\mathbf{G}_{11}^r-\mathbf{G}_{11}^a=\mathbf{G}_{11}^r(\bm{\Sigma}_0^r-\bm{\Sigma}_0^a)\mathbf{G}_{11}^a$, along with $\bm{\Sigma}_{0}^r-\bm{\Sigma}_{0}^a=-i\bm{\Gamma}_{0} $ and $\bm{\Sigma}_{0}^<=-\mathbf{f}(\bm{\Sigma}_{0}^r-\bm{\Sigma}_{0}^a)= i \mathbf{f} \bm{\Gamma}_{0}$.
the currents $I_{e}$ is finally reduced to
\begin{align}
    I_c&=\frac{e}{h} \int d\omega \mathrm{Tr}\Gamma_c[-\bm{\Gamma}_{0} \mathbf{G}_{11}^r \mathbf{f} \bm{\Gamma}_{0} \mathbf{G}_{11}^a  + \mathbf{f}\bm{\Gamma}_{0} \mathbf{G}_{11}^r \bm{\Gamma}_{0} \mathbf{G}_{11}^a ]   \nonumber\\ 
	&=\frac{e}{h} \int d\omega \mathrm{Tr}[-\bm{\Gamma}_{0,e} \mathbf{G}_{ee}^r f_e \bm{\Gamma}_{0,e} \mathbf{G}_{ee}^a -\bm{\Gamma}_{0,e} \mathbf{G}_{eh}^r f_h \bm{\Gamma}_{0,h} \mathbf{G}_{he}^a + f_e \bm{\Gamma}_{0,e} \mathbf{G}_{ee}^r \bm{\Gamma}_{0,e} \mathbf{G}_{ee}^a+ f_e \bm{\Gamma}_{0,e} \mathbf{G}_{eh}^r \bm{\Gamma}_{0,h} \mathbf{G}_{he}^a ] \nonumber \\ 
	&=\frac{e}{h} \int d\omega [f_e(\omega)-f_h(\omega)] \mathrm{Tr} [ \bm{\Gamma}_{0,e} \mathbf{G}_{eh}^r \bm{\Gamma}_{0,h} \mathbf{G}_{he}^a]\nonumber\\ 
	&=\frac{e}{h} \int d\omega [f(\omega-eV)-f(\omega+eV)] \mathrm{Tr} [ \bm{\Gamma}_{0,e} \mathbf{G}_{eh}^r \bm{\Gamma}_{0,h} \mathbf{G}_{he}^a]   , \label{S23}
\end{align}
where $T_A(\omega)=\mathrm{Tr}[\bm{\Gamma}_{0,e} \mathbf{G}_{eh}^r \bm{\Gamma}_{0,h} \mathbf{G}_{he}^a]$ is the Andreev reflection coefficient \cite{cheng_Controllable_2009, sun_Quantum_2009},
where $e,h$ in $\mathbf{G}_{ee/eh/he/hh}^{r,a}$ and $\bm{\Gamma}_{0,e/h}$ represent the electron or hole part in $\mathbf{G}_{11}^{r,a}$ and $\bm{\Gamma}_{0}$.
The Green's function $\mathbf{G}^r(E)=[\mathbf{G}^a(E)]^\dagger=\{E \mathbf{I}-\mathbf{H_{nw}}-\bm{\Sigma}^r \}^{-1}$. The differential conductance $G= dI/dV$ in the low-temperature limit is $G=\frac{e^2}{h} [T_A(eV)+T_A(-eV)]$. In calculation, we set $e=1$.

\section{\textbf{\textit{Z}} topological invariants} \label{sec100}

\begin{figure}
	\includegraphics[width=\columnwidth]{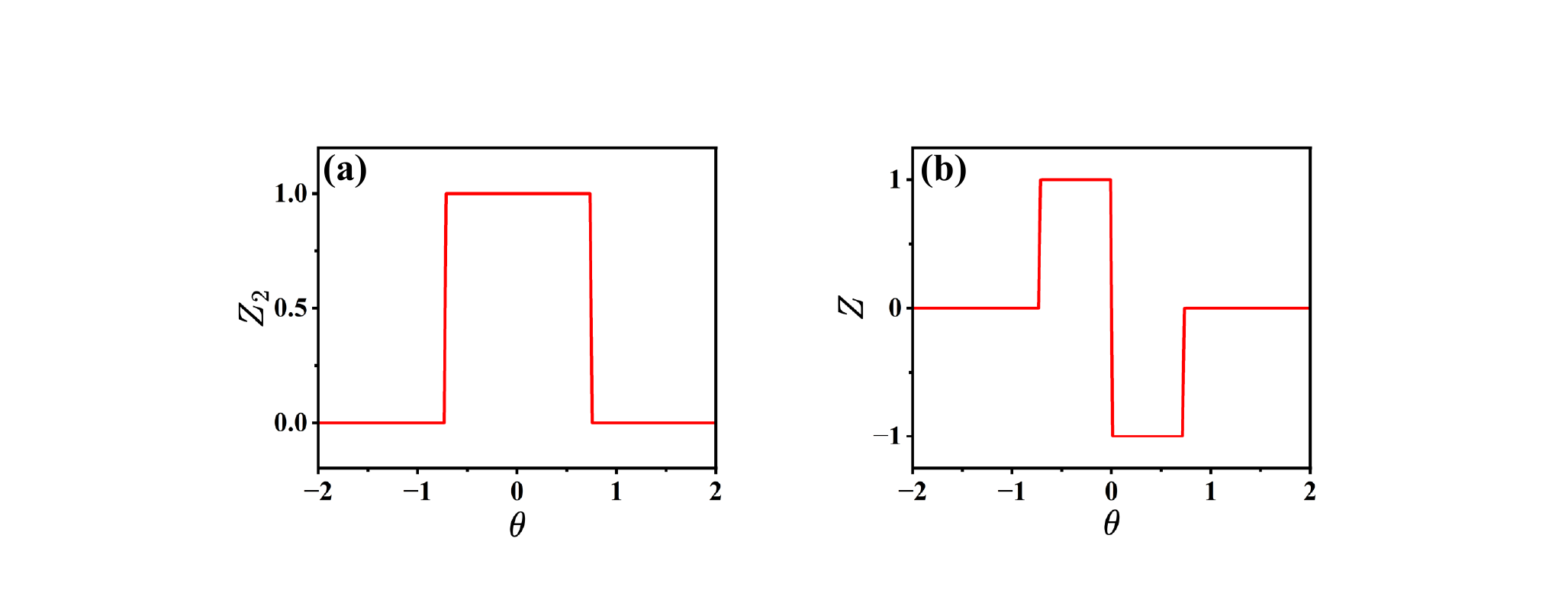}
	\caption{\label{FIGS100}
	\baselineskip 14pt
\sl{  (a) $Z_2$ number of the bulk Hamiltonian.
(b) $Z$ number of the bulk Hamiltonian.
The parameters are all the same as those in Fig. 2(c) in the main text.
}}
\end{figure}

To demonstrate the limitation of $Z_2$ invariants in capturing the phase transitions induced by chirality inversion, we also calculate the $Z$ invariants \cite{tewari_Topological_2012,
schecter_Selforganized_2016}.
As shown in Fig. \ref{FIGS100}, the region where $Z_2=1$ can be further divided into two distinct regions depending on whether $\theta \in(-\pi,0)$ or $\theta \in(0,\pi)$, corresponding to different $Z$-valued topological invariants.
Thus, at the domain wall between nanowires with opposite chirality, two MBSs are protected by chiral symmetry, leading to a $4e^2/h$ tunneling conductance as shown in the main text.

We present the derivation as follows.
In the main text, we define a chiral operator $C=\sigma_1\tau_2$, which anticommutes with the Hamiltonian $\mathscr{H}$ and $\mathscr{H}(k)$ shown in Eq. (3) and Eq. (4) in the main text, respectively.
This means that $\mathscr{H}(k)$ can be off-diagonalized in
the eigenbasis of $C$. This can be achieved by the unitary transformation $U=e^{\frac{i \pi}{4}\sigma_1 \tau_1}$, which satisfies $U^\dagger C U=diag(1,1,-1,-1)$. In
the eigenbasis of $C$, the Hamiltonian is given by

\begin{align}
    U^\dagger \mathscr{H}(k) U&=\begin{bmatrix}
  0 &\mathbf{A}  \\
  \mathbf{A}^\dagger  &  0  
\end{bmatrix}
	, \label{S200}
\end{align}
\begin{align}
    \mathbf{A}&=\begin{bmatrix}
  i(2t_J-2t_J \cos k) &\Delta+i[-2t\cos(k-\theta/2)-\mu]  \\
  -\Delta+i[-2t\cos(k+\theta/2)-\mu]  &  i(2t_J-2t_J \cos k)  
\end{bmatrix}
	. \label{S201}
\end{align}
We obtain the $Z$ value as
\begin{align}
    Z=\frac{1}{2\pi i}\oint dk \partial_k ln z(k) 
	, \label{S202}
\end{align}
where $z(k)$=det($\mathbf{A}$)/$|$det($\mathbf{A}$)$|$.

\section{\label{sec7}Calculation details of STM scanning}
The STM can be precisely moved to measure conductance of each site as shown in Fig. 3(b) in the main text. The Hamiltonian of STM is set the same as metal lead in Sec. \ref{sec6}. We consider the coupling between STM and nanowire to be weak and also set $t_M=0.2$.
In our calculations, the 1D nanowire Hamiltonian now contains two helical altermagnet nanowires with opposite chirality, each length $N=200$. The coupling between them is set to be $diag(-t,-t,t,t)$, where $t$ has been set to be $1$ in the main text.

\begin{figure}
	\includegraphics[width=\columnwidth]{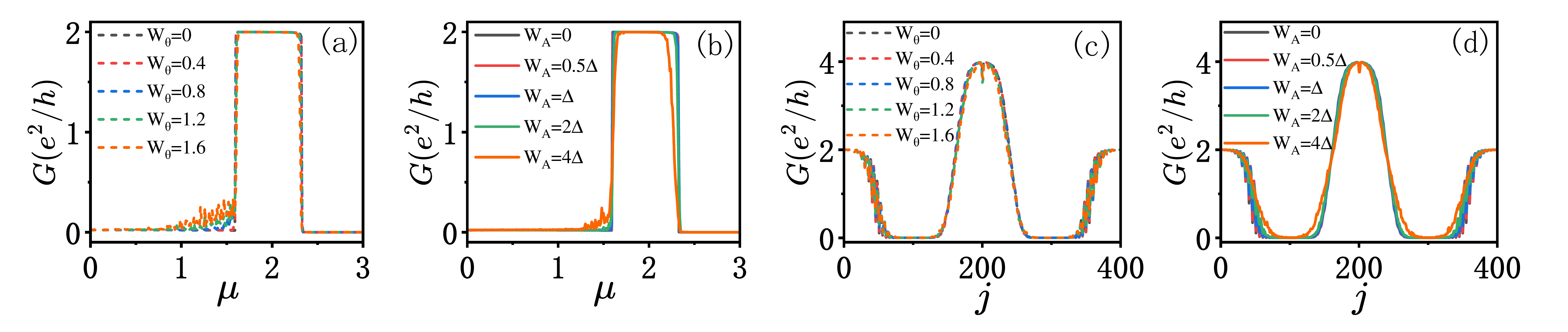}
	\caption{\label{FIGS5}
	\baselineskip 14pt
\sl{   (a-b) Calculated conductance versus chemical potential $\mu$ at zero bias for the device in Fig. 3(a) in the main text.
(c-d) Calculated conductance versus site $j$ at zero bias for the device in Fig. 3(b) in the main text.
(a,c) are with different angle disorder. (b,d) are with different Anderson disorder.
Parameters: $V=0$ in (a-d).
In (a,b), the other parameters are the same as those in Fig. 3(c) in the main text.
In (c,d), the other parameters are the same as those in Fig. 3(d) in the main text.
}}
\end{figure}

\section{\label{sec8}Impacts of defect and disorder on transport properties}
In real materials, the presence of defects inevitably induces disorder. In this section, we systematically investigate the effects of disorder.

Firstly, the disorder may disrupt the helical order.
In our theoretical model, the Néel vector orientation angle at site $j$ is denoted by the angle $\theta j $ and the hopping between sites $j$ and $j+1$ is given by $\theta (j+1/2)$ (see Eq. (2) in the main text). 
In numerical calculations, the helical order is constructed by progressively accumulating angles with a fixed step size of $\theta/2$ between adjacent hopping terms and on-site terms: 
Specifically, beginning with angle $\theta_1$ in the on-site term of initial site 1, the angle $\theta_{1\rightarrow 2}$ in hopping term between site 1 and site 2 is generated via $\theta_1+\theta/2$, the angle $\theta_2$ in the on-site term of site 2 is generated via $\theta_{1\rightarrow 2}+\theta/2$, with this iterative pattern generating the helical order.
To simulate helical order disorder (angle disorder), we add the disorder term $\delta\theta_i$ to the original step size $\theta/2$, where $\delta\theta_i\in [-W_{\theta}/2, W_{\theta}/2]$ follows a uniform distribution and $W_{\theta}$ is the angle disorder strength.
In Figs. 3(c) and (d) of the main text, we show the conductance versus chemical potential $\mu$ at nanowire ends and conductance versus STM position $j$ along nanowires with opposite chirality, respectively. Based on these two configurations, we investigate the zero-bias conductance 
under different angle disorder strengths in Figs. \ref{FIGS5}(a) and (c).
Both the $2e^2/h$ zero-bias peak at nanowire ends in Fig. \ref{FIGS5}(a) and the $4e^2/h$ STM tunneling conductance at the domain wall of nanowires with opposite chirality in Fig. \ref{FIGS5}(c) remain robust against the disorder-induced disruption of the helical order.

Secondly, we consider the Anderson disorder in our system and introduce a disorder chemical potential of site $j$ as $\mu_j=\mu+\delta\mu_j$, and $\delta\mu_j$ is uniformly distributed in the range $[-W_A/2, W_A/2]$ with Anderson disorder strength $W_A$.
Based on the two configurations in Figs. 3(c) and (d) of the main text, we also investigate the zero-bias conductance 
under different Anderson disorder strengths in Figs. \ref{FIGS5}(b) and (d).
The topological superconducting gaps are generally smaller than the proximity-induced gap $\Delta$, as demonstrated in Figs. \ref{FIGS1}(d) and (e).
In calculation, we choose five $W_A$ values from $0$ to $4\Delta$, where $4\Delta$ is a large value compared with topological superconducting gap.
Figure \ref{FIGS5}(b) illustrates the zero-bias conductance versus chemical potential $\mu$ for different $W_{A}$.
When $W_{A}=0$, the $2e^2/h$ conductance plateau exists in the range about $1.6<\mu<2.3$.
As $W_A$ increases further, the plateau remains almost unchanged.
Figure \ref{FIGS5}(d) illustrates STM tunneling conductance versus site $j$ for different $W_{A}$. 
The $4e^2/h$ conductance at the domain wall persists even for the large $W_A$.

In conclusion, even when the helical order is partially disrupted or the large Anderson disorder exists, the MBSs remain localized at ends of helical altermagnet nanowire, with the transport calculations exhibiting a zero-bias conductance peak of $2e^2/h$.
Furthermore, the $4e^2/h$ zero-bias peak at the domain wall of nanowires with opposite chirality remains robust against the defect and disorder.

\bibliography{suprefer.bib}